# Structure of Molten $FeCl_2$ and $FeCl_3$

Fakhrul Hasan Bhuiyan[1], Jicheng Guo[2], Christopher James Benmore[3], Avery Blockmon[2], Denis Johnson[2], Alvaro Vazquez Mayagoitia[1]

[1]Computational Science Division, Argonne National Laboratory

[2]Chemical and Fuel Cycle Technologies, Argonne National Laboratory

[3]X-ray Science Division, Argonne National Laboratory

* Fakhrul Hasan Bhuiyan, Jicheng Guo

**Email:** fbhuiyan@anl.gov, jicheng.guo@anl.gov

## Abstract

Molten iron chlorides are central to emerging energy technologies, including electrochemical iron production and redox flow batteries. Optimizing their electrochemical performance and transport properties requires atomic-scale structural understanding, yet detailed data for molten $FeCl_2$ and its differences from $FeCl_3$ remain scarce. Here, we determined the structures of molten $FeCl_2$ and $FeCl_3$ using High Energy X-ray diffraction (HEXRD), Empirical Potential Structure Refinement (EPSR), and molecular dynamics (MD) simulations with machine learning interatomic potentials (MLIPs). HEXRD measurements provided structure factors and total radial distribution functions (RDFs), which were quantitatively reproduced through EPSR refinement directly constrained by experimental data. MD simulations using MACE foundation and fine-tuned models reproduced experimental structure factors as well as total and partial RDFs, capturing key structural differences between the melts. The models resolved the octahedral to tetrahedral coordination transition of Fe upon melting in $FeCl_3$ and predicted a similar transition in $FeCl_2$. Analysis of MD trajectories quantified coordination environments, bridging Cl populations, bond-angle distributions, and connectivity patterns, revealing distinct degrees of polymerization and local geometry. Polymer chain statistics further showed that, contrary to prior reports, both liquids predominantly consist of extended chains containing six or more Fe centers rather than discrete $Fe_2Cl_6$ units. Finally, diffusion coefficients of the two melts calculated from the MACE-MD simulations were compared. Together, these results establish atomic-scale structural benchmarks for molten $FeCl_2$ and $FeCl_3$ and demonstrate the reliability of MACE-based MLIPs for predictive modeling of high-temperature molten salts, while providing practical guidance for MLIP development in complex ionic liquids.

## Introduction

The structural properties of iron chlorides in both crystalline and liquid states are fundamental to understanding the chemistry of iron-based systems and their applications in industrial processes, including catalysis, metal refining, electrochemical processing, and molten salt technologies.[1–4] Iron exists in two primary oxidation states in iron chlorides: Fe(II) in iron(II) chloride ($FeCl_2$) and Fe(III) in iron(III) chloride ($FeCl_3$), each associated with distinct coordination geometries and electronic properties.[5,6] In the crystalline state, $FeCl_2$ adopts the $CdCl_2$-type layered structure with trigonal symmetry, where $Fe^{2+}$ ions occupy octahedral sites coordinated by six chloride ions.[2] Similarly, crystalline $FeCl_3$ exhibits the $BiI_3$-type (Bismuth tri-iodide) structure, featuring octahedral coordination consistent with six-fold coordination symmetry.[2,7] Melting disrupts these ordered frameworks and can produce liquid structures that differ substantially from their crystal structure, making structural investigation of the liquid state essential for both fundamental understanding and practical applications.

Upon melting, the local structure of iron chlorides undergoes dramatic reorganization which is associated with a 34% decrease in density of $FeCl_3$.[8–10] Neutron diffraction measurements showed that melting of $FeCl_3$ was accompanied by a marked change in local coordination, with $Fe^{3+}$ shifting from six-fold octahedral coordination in the crystal to predominantly four-fold tetrahedral environments in the liquid.[2] Further investigations combining neutron diffraction and X-ray diffraction studies showed that molten $FeCl_3$ could be described in terms of closely packed $Fe_2Cl_6$ bi-tetrahedral or molecular units, similar to those observed in the vapor phase, but with strong intermolecular interactions in the condensed phase.[2,7] The measured paramagnetic moments of molten $FeCl_3$ were approximately half of those expected for free $Fe^{3+}$ ions, which suggested partial charge transfer and intermediate ionic character in the melt.[7] Similar octahedral-to-tetrahedral coordination changes on melting were reported for $AlCl_3$.[2,11] These observations established that melting in iron chlorides can involve major rearrangement of the local coordination environment rather than simple retention of crystalline order. Such structural reorganization can have profound implications for ionic conductivity, viscosity, and other transport properties that govern industrial processes.[11–13]

Related studies of iron oxide melts provide additional experimental context for understanding such coordination changes in iron chloride melts. High-energy X-ray diffraction (HEXRD) measurements of molten iron oxides across the FeO to $Fe_2O_3$ compositional range showed that average Fe-O coordination numbers were approximately 4.5 to 5, which was substantially lower than the six-fold octahedral coordination observed in crystalline FeO and $Fe_2O_3$.[1,14] The structural complexity of these melts, including mixed oxidation states and variable coordination geometries, reinforced the broader view that melting can strongly reorganize local environments in iron-containing systems. Although experimental techniques have provided critical structural information, obtaining a complete atomic-scale picture of such complex liquids from experiments alone remains difficult. Computational tools are therefore often used to complement experimental findings.

Empirical potential structure refinement (EPSR) is a complementary modeling technique that uses Monte Carlo simulation constrained by experimental diffraction data to refine atomistic structural models.[15,16] EPSR operates by using diffraction data to perturb reference potentials, derived from literature or constructed from first principles, until simulated structure factors converge with measured experimental data. The method employs harmonic force constants to define molecular geometries and Lennard-Jones potentials to describe interatomic interactions, allowing incorporation of chemical constraints and effective charges. EPSR has proven effective for molten salt systems, capturing both short-range order and intermediate-range order and refining atomistic structural models for iron-containing melts.[1,14]

First-principles molecular dynamics (MD) simulations based on density functional theory (DFT) have become standard tools for probing liquid structure at the atomic scale. Such DFT-based simulations can reproduce experimentally measured structural observables while also providing dynamical information and local coordination statistics that are not directly accessible from diffraction alone. In molten salts, first-principles MD has successfully reproduced total structure factors, pair distribution functions, and vibrational signatures, supporting X-ray scattering and Raman spectra.[17,18] For example, molten $MgCl_2$ and $ZnCl_2$ salts have been studied using ab initio MD which resolved the populations of $MgCl_5$ and $ZnCl_4$ units and linked spectral features to specific polyhedral connectivities.[17] For Fe-containing chlorides, ab initio calculations were used to complement Raman spectroscopy and predict the vibrational frequencies of $FeCl_3$, $[FeCl_4]^-$, and $[Fe_2Cl_7]^-$ in $FeCl_3$/$FeCl_2$-based ionic liquids.[19] These capabilities make first-principles MD particularly valuable for probing coordination geometries and local bonding arrangements in complex iron chloride liquids. The high computational cost of these calculations, however, limits the accessible time and length scales.

Recent developments in machine learning have enabled construction of interatomic potentials that bridge the gap between the high accuracy of density functional theory and the efficiency of empirical force fields.[20–23] Machine learning interatomic potentials (MLIPs) are trained on quantum-mechanical reference data to predict atomic energies and forces, enabling MD simulations with near-DFT accuracy at much lower computational cost. The resulting increase in accessible system size and simulation time makes these approaches particularly attractive for molten salt systems, where intermediate-range order and structural fluctuations may require broader spatial and temporal sampling.

The Gaussian Approximation Potential (GAP) framework, originally proposed for interpolating quantum mechanical potential energy surfaces using Gaussian process regression, represents one of the foundational approaches in this field.[21,22] GAP models have been successfully applied to various materials systems, including bulk silicon, carbon, tungsten, and iron, demonstrating the broad applicability of this methodology.[21,22] Unfortunately, GAP models face some computational challenges, in particular the Smooth Overlap of Atomic Position (SOAP) descriptor, often are limited by the steep growth of the descriptor size, large memory footprint, and lack of support of GPU (graphics processing units). On the other hand, neural network (NN) potentials have emerged as another powerful approach for constructing high-dimensional potential energy surfaces from ab initio data.[23] These methods construct the total energy as a sum of environment-dependent atomic energies, enabling application to high-dimensional systems containing thousands of atoms. Such NN potentials have been applied to molten salt systems, including alkali chlorides such as NaCl and LiCl, demonstrating excellent agreement with both DFT calculations and experimental data in analyzing structural properties, chemical potential, vibrational spectra.[24–27]

Foundation models represent an emerging paradigm in MLIPs, where models are trained on large, diverse datasets spanning multiple chemical systems and phases, enabling transfer learning and zero-shot predictions on unseen systems.[28,29] Specifically, the MACE framework exemplified this approach by employing higher-order equivariant message passing neural networks for fast and accurate force field construction.[30–32] Notably, MACE models require less training data than GAP and other non-equivariant NN potentials to reach similar predictive performance. The MACE-MP-0 foundation model, trained on inorganic crystal structures from the Materials Project database, demonstrated remarkable capability for running stable MD for a wide range of molecules and materials.[30] The foundation model approach offers the potential to capture physics relevant to multiple phases (crystalline and liquid) and multiple oxidation states without system-specific model fitting, though performance on systems significantly different from the training distribution requires careful validation. Although applications of MACE-based foundation models to molten salts, and especially to iron-containing salts, remain limited, a recent study has demonstrated that the MACE-framework can be utilized to model diverse molten chloride systems with near-DFT-accuracy and reproduce key thermophysical and structural properties.[33] These results highlight the promise of foundation-model-based approaches for extending atomistic simulations of complex molten salts to broader composition, length, and time scales.

Despite substantial progress in experimental characterization and computational modeling, important questions remain regarding the structures of $FeCl_2$ and $FeCl_3$ melts. In particular, the structure of molten $FeCl_2$ and its differences from molten $FeCl_3$ remain poorly understood. Moreover, although neutron and X-ray diffraction studies have demonstrated that the coordination environment of Fe in $FeCl_3$ changes upon melting, a clear atomic-scale description of the local structure in $FeCl_3$ liquid is still lacking. Foundation MLIP models remain largely unexplored for molten iron chloride systems despite the potential utility of the foundation models. Therefore, validation against experimental diffraction data from both crystalline and liquid phases is essential to establish the reliability of foundation MLIP models for molten salts. In this study, these gaps were addressed through a comprehensive multi-method investigation combining HEXRD measurements, EPSR calculations, and MD simulations using MACE foundation and finetuned models of liquid $FeCl_2$ and $FeCl_3$. The combination of advanced experiments with complementary computational approaches provided a robust framework for resolving the atomic-scale structure of the molten salt systems and for evaluating the accuracy and transferability of MACE models in describing phase transitions and coordination changes in molten ionic materials.

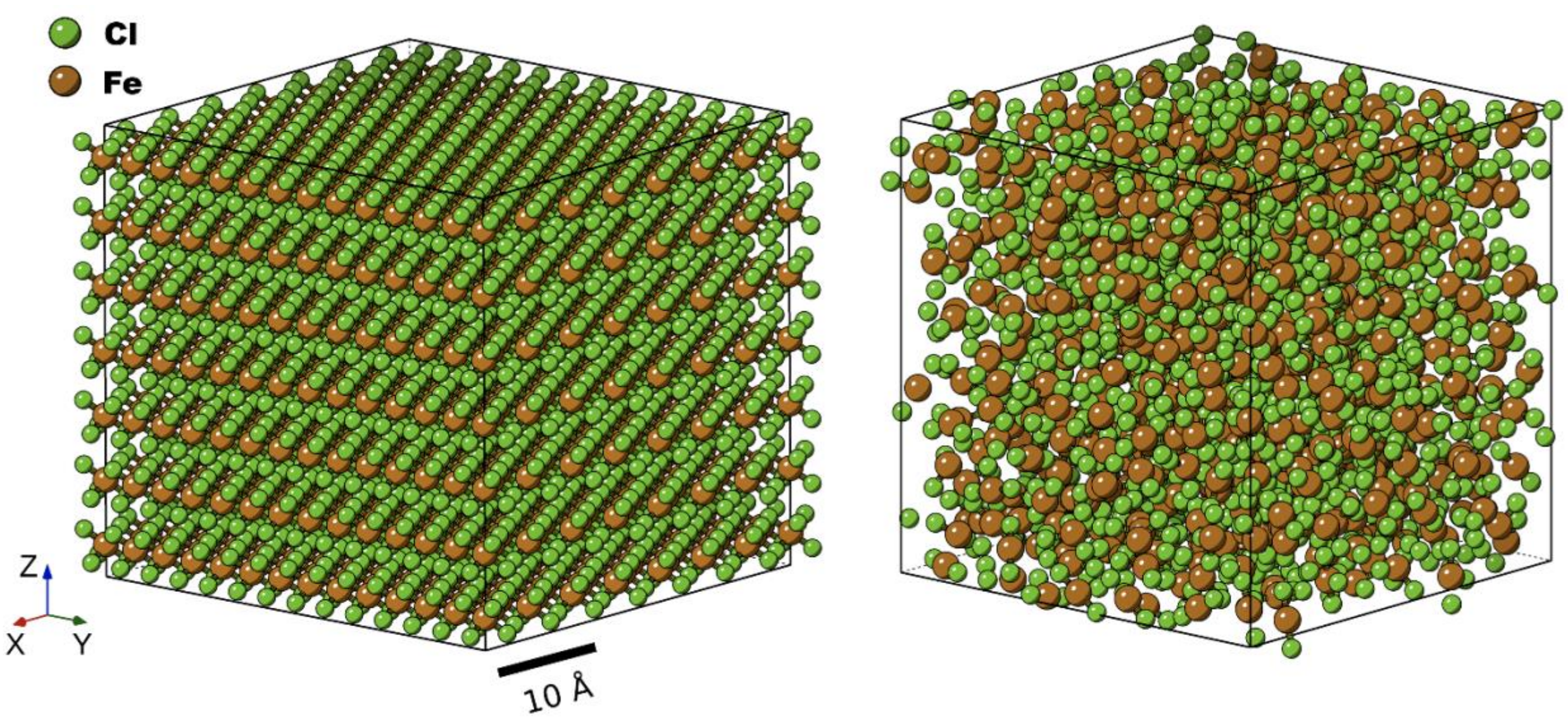

**Figure 1:** $FeCl_2$ crystal (left panel) and liquid (right panel) systems for MD simulations. Green atoms represent Cl and brown atoms represent Fe.

## Methods

### Experiments

$FeCl_2$ (Sigma-Alrich, ≥ 99.99 trace metal basis) and $FeCl_3$ (Sigma-Aldrich,≥ 99.99 trace metal basis) were loaded in separated fused silica quartz tubes (VitroCom) in a UHP Ar glovebox (<0.1 ppm $O_2$, < 1ppm $H_2O$), which were then sealed under vacuum after being taken out from the container. HEXRD measurements on liquid and crystalline $FeCl_2$ and $FeCl_3$ were conducted at beamline 6-ID-D of the APS in transmission geometry, using an incident energy of 90.1 keV. The wide-angle X-ray scattering (WAXS) detector diffraction patterns were calibrated using a NIST $CeO_2$ standard, and analyzed using Fit2D software (Hammersley 2016), which corrects for flat field, polarization, detector rotation and tilt. For liquid salt structure measurement, the salt sample in the quartz tube was heated to the melting temperature of the salt (602°C for $FeCl_2$ and 304°C for $FeCl_3$) until all the crystalline signals disappeared. And the diffraction pattern was recorded for 300s for each temperature for better statistics.

The program PDFgetX2[34] was used to correct background, oblique incidence, absorption and detector efficiency effects and normalize the I(Q) data to the sample self-scattering in absolute electron units. The extracted total x-ray structure factor, S(Q) was truncated at a maximum Q-value, corresponding to a positive node at Q=20 $Å^{-1}$, and Fourier transformed using a Lorch modification function to minimize truncation ripples and yield the differential distribution function, D(r). Here we use the formalism of Hannon-Howell-Soper to describe the RDF.[35]

**Simulation Details**

MD simulations were performed using the LAMMPS[36] simulation package with the MACE[31] machine-learning interatomic potential. MACE force fields were deployed in LAMMPS using the Symmetrix[30,32] framework. Three MACE foundation models and one fine-tuned MACE model were employed to simulate both crystalline and molten phases of $FeCl_2$ and $FeCl_3$. Crystalline systems were simulated at 300 K. Molten $FeCl_2$ was simulated at 950 K, 1050 K, and 1150 K, while molten $FeCl_3$ was simulated at 650 K, consistent with experimental conditions.

All simulations were conducted in the canonical (NVT) ensemble using a cubic simulation cell of approximately 4 nm per side. A time step of 1 fs was used, and each simulation was run for 100 ps. All systems were observed to reach equilibrium well within the simulation time. Structural analyses, including radial distribution functions (RDFs) and structure factors, were computed using configurations collected from the final 10 ps of each trajectory. MD simulations to calculate diffusion coefficient in molten $FeCl_2$ and $FeCl_3$ were run for 1 ns of total simulation time. Simulation systems were created with the help of Atomsk[37] and Packmol[38] packages. Visualization was performed using OVITO[39] and CrystalViewer[40] software. All simulations and fine-tuning process was orchestrated and executed through ParslBox which is a Python package for workflow orchestration and execution.

**Fine-tuning of MACE Model**

Fine-tuned MACE model was obtained by fine-tuning the MACE-$r^2$SCAN foundation model using a dataset generated from DFT calculations. Training structures were selected by first sampling configurations from MD trajectories, followed by DFT labeling. Sampling was performed for $FeCl_2$ crystal structures at 300 K and 500 K, and for $FeCl_2$ liquid structures at 950 K, 1050 K, and 1150 K, using appropriate experimental densities. Each sampling trajectory was generated from a 100 ps NVT simulation.

From each trajectory, representative snapshots were selected using a furthest point sampling strategy based on the SOAP descriptors.[41,42] Simulation box dimensions were limited to below 15 Å for all sampling simulations. A total of 50 configurations were selected from each trajectory. For each sampled configuration, energies, atomic forces, and virial stresses were computed using DFT with the $r^2$SCAN meta-generalized gradient approximation exchange-correlation functional.[43] A plane-wave energy cutoff of 700 eV was employed together with a k-point spacing of 0.25 $Å^{-1}$. Electronic occupancies were treated using Methfessel-Paxton smearing of order 2 with a smearing width of 0.1 eV. All calculations were spin-polarized. The electronic self-consistency criterion was set to $10^{-7}$ eV. All calculations were performed as single-point evaluations on the sampled structures using the Vienna Ab initio Simulation Package.[44]

Only structures that reached self-consistent field convergence were retained, and unconverged calculations were discarded. The converged structures were further characterized based on the maximum per-atom force and energy drift, i.e., structures exhibiting large forces or energy drift were removed. Specifically, configurations with per-atom forces exceeding 10 eV/Å or energy drift greater than 0.1 eV were excluded from the training dataset. Finally, the constructed training dataset was used to fine-tune the MACE-$r^2$SCAN foundation model. Multi-headed training was employed where the training dataset was augmented with additional structures from the original foundation model training dataset. All original model hyperparameters were retained during fine-tuning. A summary of training performance metrics is provided in Table 1.

**Table 1:** Evaluation of the fine-tuned MACE model using the root mean squared error (RMSE) in energy and force

| Dataset | RMSE Energy (meV/atom) | RMSE Force (meV/Å) |
|---|---|---|
| Train | 2.1 | 56.9 |
| Validation | 2.6 | 99.0 |

## Results

### Experimental Findings

Figure 2 shows the structure factor, S(Q), and the RDF, G(r), of molten $FeCl_2$ and $FeCl_3$ obtained from the HEXRD experiments. In S(Q), the first two peak positions between Q = 1-3 $Å^{-1}$ coincided for both melts, although the peak heights were lower for $FeCl_2$. This indicates that the periodicity of the intermediate-range structure is similar in the two melts, while the reduced peak intensity in $FeCl_2$ is consistent with a more disordered arrangement. At $Q > 3$ $Å^{-1}$, the peak positions for $FeCl_2$ and $FeCl_3$ began to deviate from each other, while the peak heights remained comparable.

The RDF plots reveal distinct difference in the short-range structures between the two melts, which reflected in the distance of the atoms within first coordination shell. The first peak, which is mainly from the Fe-Cl interaction, for $FeCl_2$ was centered at 2.33 Å, while that for $FeCl_3$ appeared at 2.19 Å, consistent with the shorter Fe–Cl bond length expected for the higher oxidation state of $Fe^{3+}$. The position of the second peak which corresponds mainly to the Cl-Cl interaction were largely similar, at 3.65 Å for $FeCl_2$ and 3.61 Å for $FeCl_3$, although the peak height was greater for $FeCl_3$, likely due to the higher amount of Cl-Cl interaction in $FeCl_3$. It should be noted that the RDF values below 2 Å represent an unphysical region, which provides an estimate of the level of error present in the data and deviations from the spherical atom approximation.

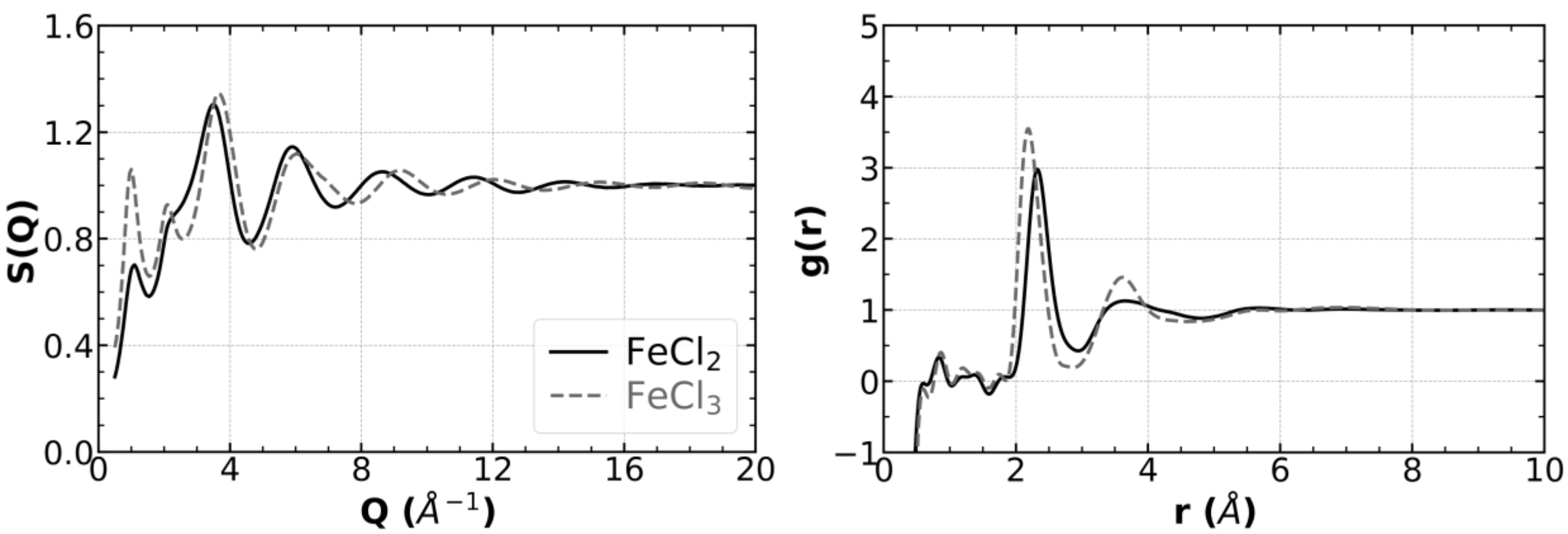

**Figure 2:** HEXRD-derived structure factors (left panel) and total RDFs (right panel) for molten $FeCl_2$ (750°C) and $FeCl_3$ (310°C).

The molten structure of $FeCl_3$ was previously studied using neutron diffraction, where the first and second peak positions were reported at 2.22 Å and approximately 3.7 Å, respectively.[2,7] The first peak position observed in our HEXRD measurement (2.19 Å) matches closely with the neutron diffraction result, confirming the Fe–Cl nearest-neighbor distance. The second peak has been predominantly attributed to Cl–Cl interactions based on electron diffraction data on the $Fe_2Cl_6$ molecule in the vapor phase, which was fitted by either a planar ($D_{2h}$) or puckered ($C_{2v}$) bitetrahedral configuration.[45,46] The second peak position observed here (3.61 Å) is in close agreement with the Cl–Cl distance reported for both the planar ($D_{2h}$) and puckered ($C_{2v}$) bitetrahedral configurations (3.59 Å). Overall, the present HEXRD results and the earlier neutron diffraction measurements yielded consistent short-range and intermediate-range structural features and ordering in the $FeCl_3$ melt.

**Computational Findings**

To investigate the structure of the melts more closely, MD simulations using three different MACE foundation models, namely, MACE-MPA-0, MACE-MATPES-PBE-0, and MACE-MATPES-$r^2$SCAN-0, were conducted. Comparison of all MACE model-calculated structure factors and RDFs with experimental results are shown in Figure S1 in the supplementary information (SI). All structure factors and RDFs reported from simulations in this text were averaged over the last 10 ps of corresponding simulation trajectory. The MPA-0 and MATPES-PBE-0 MACE foundation models overpredicted the first peak intensity in the $FeCl_2$ structure factor and produced a flatter second peak for $FeCl_3$ relative to experimental structure factors. In addition, both models shifted the first peak of $FeCl_2$ to lower Q values. In contrast, the MACE-MATPES-$r^2$SCAN-0, hereafter referred to as MACE-$r^2$SCAN, showed the best overall agreement with the experimental structure factors and RDFs for both $FeCl_2$ and $FeCl_3$. In the following text here, results from only the MACE-$r^2$SCAN model and its fine-tuned version will be discussed since those were the best performing models.

Figure 3 shows the structure factor and RDF for $FeCl_2$ obtained from the MD simulations with the MACE-$r^2$SCAN foundation model. The simulated structure factor followed the experimental structure factor from the HEXRD measurements very closely for $Q > 3$ Å$^{-1}$, indicating that the short-range structure of the $FeCl_2$ melt was accurately captured by the MACE force field. For $Q < 3$ Å$^{-1}$, a slight mismatch between the simulated and experimental structure factors was observed. The first peak in the simulated structure factor was shifted to a lower Q value (0.94 Å$^{-1}$) compared to the experimental peak (1.08 Å$^{-1}$), suggesting slight discrepancies between simulated and actual intermediate-range structural ordering. Additionally, the simulations, with a slightly higher first peak height, predicted a higher degree of intermediate-range ordering than what was observed experimentally. To improve agreement with experiment, the MACE-$r^2$SCAN foundation model was further fine-tuned using additional crystal and liquid $FeCl_2$ structures. The structure factor obtained from MD simulations with the fine-tuned model closely matched that of the original foundation model. Although the position of the first peak did not shift from the foundation model calculated peak position, the peak intensity decreased, resulting in better agreement with the experimental peak height relative to the foundation model. The small overestimation of the first peak height in simulations could partly reflect finite-size effects associated with the limited simulation box size, which can affect the sampling of density fluctuations and, in turn, the low-Q and peak-amplitude behavior.

Similar discrepancies in the intermediate-range ordering region (low-Q) have previously been reported for $MgCl_2$ and $ZnCl_2$ melts.[17] In those studies, fist-principle MD simulations employing various exchange–correlation functionals, including PBE-D3, PBE0-D3, and SCAN-D3, consistently showed slight deviations from scattering experiments in the low-Q region. Taken together, these comparisons suggest that the observed discrepancy in the intermediate-range structure between experiments and simulations is unlikely to originate from the MACE framework itself. Rather, achieving full quantitative agreement with experiment in the intermediate-range regime may require an exchange–correlation functional better suited than $r^2$SCAN for describing this system.

To further assess the structure of the molten $FeCl_2$, the HEXRD data were also analyzed using EPSR where an empirical interaction potential was iteratively adjusted to reproduce the measured scattering data while maintaining chemically reasonable local environments. The resulting EPSR model reproduced the experimental structure factor more closely in the low-Q region (Figure 3), demonstrating that a chemically consistent structural model was able to match the measured intermediate-range correlations. In case of the RDF, both the MACE-MD simulations and the EPSR modeling reproduced the experimental total RDF closely (Figure 3).

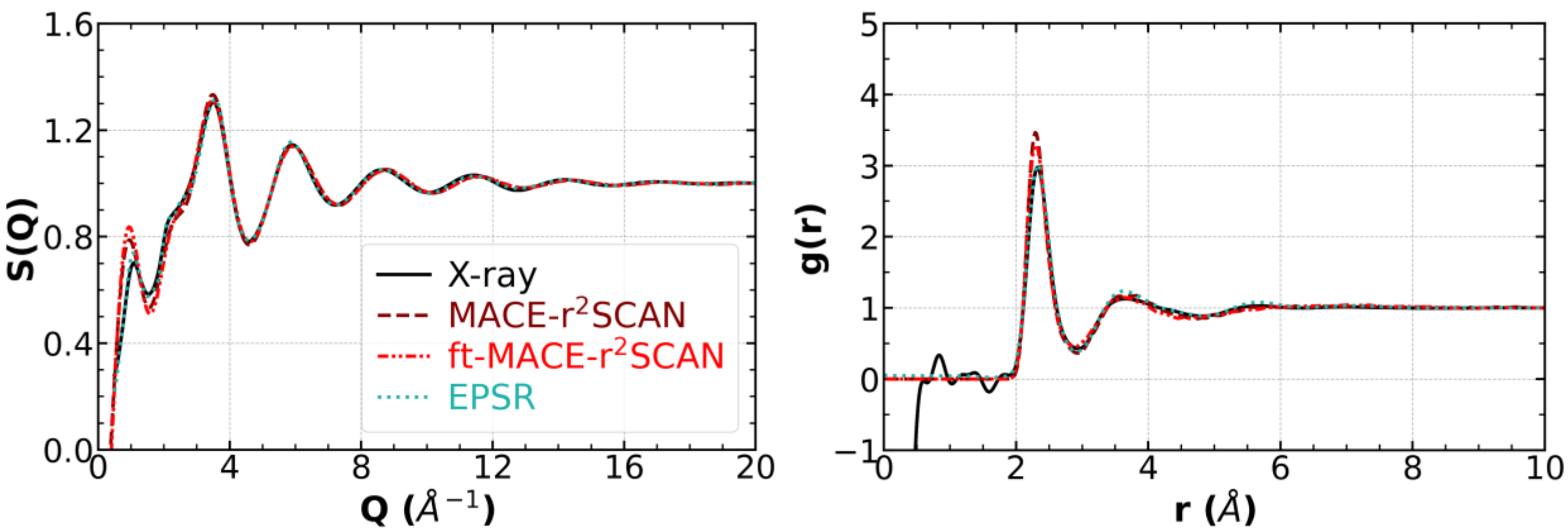

**Figure 3:** Comparison of experimental structure factor (left panel) and RDF (right panel) for $FeCl_2$ with results from MD simulations with MACE foundation model (MACE-r$^2$SCAN) and finetuned model (ft-MACE-r$^2$SCAN), and EPSR modelling.

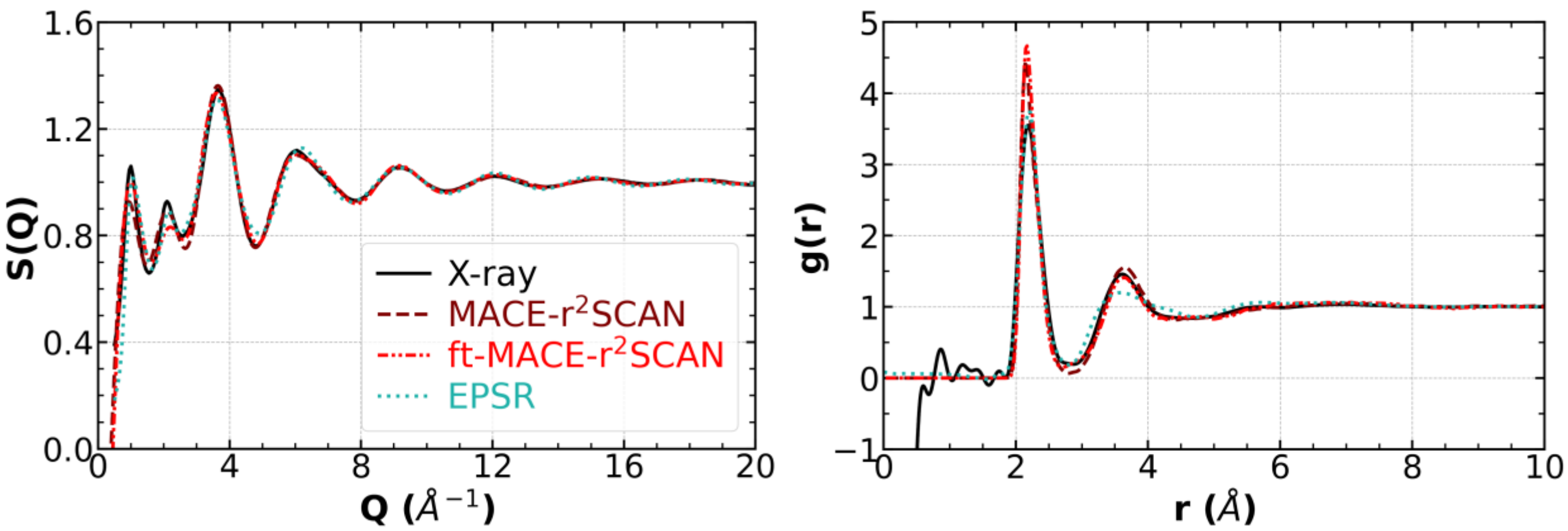

**Figure 4:** Comparison of experimental structure factor (left panel) and RDF (right panel) for $FeCl_3$ with results from MD simulations with MACE foundation model (MACE-r$^2$SCAN) and finetuned model (ft-MACE-r$^2$SCAN), and EPSR modelling.

The structure factor and RDF for $FeCl_3$ obtained from MD simulations with the MACE models are shown in Figure 4. Both the foundation and fine-tuned MACE models, and the EPSR model reproduced the experimental structure factor closely, with only minor differences in the heights of the first two peaks in the MD-derived structure factors. In case of the RDF, the MACE models reproduced the experimental total RDF more closely than the EPSR model. Notably, the fine-tuned model was trained using only $FeCl_2$ data; however, multi-headed fine-tuning, i.e., augmenting the training dataset with a subset of the original model training data, helped the model retain the predictive capabilities of the original foundation model.

After establishing reasonably good agreement in the total structure factor and total RDF from HEXRD experiments and MD simulations, the local structures of the two melts were examined using partial RDFs. Figure 5 shows the partial RDFs of $FeCl_2$ and $FeCl_3$ from the MD simulations and EPSR modeling. A key structural difference between the two melts is observed in the first peak position of the Fe-Cl partial: for $FeCl_2$, the peak is shifted by approximately 0.1 Å to larger distances relative to $FeCl_3$. The EPSR model further predicted a slightly shorter Cl-Cl first-peak distance for $FeCl_3$ (3.48 Å) compared to $FeCl_2$ (3.54 Å), reflecting a more compact packing of chloride ligands around $Fe^{3+}$.

The MACE models and the EPSR model deviated significantly in predicting the Fe-Fe partial RDF and the second peak of the Fe-Cl correlation. The MACE models consistently predicted a shorter and broader Fe-Fe first peak compared to the EPSR model. For $FeCl_3$, the finetuned model deviated further from the foundation model, predicting an even shorter Fe-Fe first-peak. Given the higher charge on $Fe^{3+}$, stronger cation-cation repulsion would generally be expected to shift Fe-Fe correlations to larger distances compared to $FeCl_2$ melt. The predicted shortening of the Fe-Fe peak by the fine-tuned model therefore maybe an artifact of the finetuning process, where the inclusion of additional $FeCl_2$ training data could have disproportionately influenced $Fe^{2+}$ interactions and indirectly affected the $Fe^{3+}$ description.

Despite these differences in the Fe-Fe and second-shell features, the MACE models were consistent with each other and with the EPSR model in terms of Fe-Cl coordination number, yielding a value of approximately 4 in the liquid phase. Crystal equilibration simulations were also performed for both $FeCl_2$ and $FeCl_3$ using the foundation and fine-tuned MACE models. The partial RDFs and the coordination number of crystal $FeCl_2$ and $FeCl_3$ from the MD simulations are shown in Figure S2. A comparison of the coordination numbers in the crystalline and liquid phases is shown in Figure S3. The MACE models predicted a Fe-Cl coordination number of 6 in the crystalline phases (Figure S3). These results indicated that the MACE models predicted a transition in the local coordination environment, from octahedral Fe-Cl coordination in the crystals to predominantly tetrahedral coordination in the melts. Figure 6 shows octahedral coordination of Cl ions around Fe in both $FeCl_2$ and $FeCl_3$ crystals, and tetrahedral coordination of Cl around Fe in the corresponding melts. Such structural transition from crystal to liquid accompanied by a 34% decrease in density has previously been reported for $FeCl_3$.[2,7] The present results suggest that $FeCl_2$ undergoes a similar structural transition, accompanied by a 26% decrease in density upon melting.

## Discussion

### Revisiting $FeCl_3$ Liquid State Model

The currently accepted liquid state model of $FeCl_3$ is based on the neutron and x-ray diffraction measurements of Price et al.[2] and Badyal et al.[7] who proposed the liquid comprises almost entirely of $Fe_2Cl_6$ molecules. This has been supported by Raman scattering measurements and MD simulations using polarizable ionic interaction model of molten iron (III) chloride.[47,48] To test this hypothesis, we have conducted two EPSR simulations on our liquid $FeCl_3$ diffraction data. The first model comprises solely of $Fe_2Cl_6$ molecules, and the second model starts from a random mixture of individual Fe and Cl atoms. To achieve an approximate fit to the HEXRD data for the first model, it was necessary to make the $Fe_2Cl_6$ molecules flexible. A similar misfit to the pair distribution function was reported by Price et al.[2] Although the correct 4-fold coordination environment was obtained, the $Fe_2Cl_6$ model did not fully capture the observed disorder in the first and second peaks in G(r). More importantly, to achieve the necessary intermolecular packing, mainly reflected by the first sharp diffraction peak in S(Q), the model did not reproduce the observed experimental density. This is illustrated in Figure 7 where to achieve the necessary short-range order between molecules, the simulation box could not be filled properly, indicating the $Fe_2Cl_6$ molecular structure model is associated with a much lower density fluid. In contrast, the $FeCl_3$ atomic model produced an almost exact match to both the experimental x-ray and density data but did not contain any $Fe_2Cl_6$ molecules. Rather, the liquid $FeCl_3$ atomic model showed corner-shared $FeCl_4$ polyhedra, forming short winding chains.

The composition of the $FeCl_3$ melt and the distribution of $FeCl_4$ polyhedral chain lengths were further analyzed using MD simulations. Figure 8 presents the weight fraction of polyhedral chains compared to the total simulation box weight in the liquid $FeCl_3$ simulation cell over time. Chains were identified based on Fe–Cl connectivity, using a cutoff distance of 3 Å to define an Fe-Cl bond, selected from the first minimum of the Fe-Cl partial RDF shown in Figure 5. With this criterion, $Fe_2Cl_6$ units accounted for only about 15% of the weight of the total simulation cell, whereas extended chains containing six or more Fe centers comprised approximately 70% of the simulation box weight. As expected, the quantitative distribution depends on the chosen bond cutoff. Reducing the cutoff to 2.5 Å increases the apparent fraction of $Fe_2Cl_6$ species and decreases the fraction of longer chains (Figure S4). However, even under this stricter definition, chains containing six or more Fe centers remained the dominant structural motif at roughly 30 wt%, compared to about 20 wt% for $Fe_2Cl_6$ units. Overall, both MACE-MD simulations and EPSR modeling consistently indicated that the conventional description of molten $FeCl_3$ as consisting primarily of discrete $Fe_2Cl_6$ molecules is incomplete. Instead, the liquid is dominated by extended polyhedral chain networks.

**Comparison Between $FeCl_2$ and $FeCl_3$ Melts**

While the change in Fe-Cl bond length is anticipated between liquid $FeCl_2$ and $FeCl_3$, a coordination increase with the change in oxidation state may also be expected based on existing vapor phase and crystalline structures. This is because in the vapor phase Fe atoms are 4-coordnate in the form of $Fe_2Cl_6$ dimers with bond lengths of r(Fe-Cl)=2.13Å (terminal) - 2.33Å (bridging),[49] while in the crystalline state Fe atoms are octahedrally coordinated to six chloride ions with r(Fe-Cl) = 2.27-2.51 Å.[10] Therefore, based on our observed bond lengths in the liquid state compared to those in the vapor and crystals, one might expect liquid $FeCl_2$ to exhibit sixfold Fe-Cl coordination compared to the fourfold Fe-Cl coordination in liquid $FeCl_3$. However, our results showed that this anticipated coordination environment in the $FeCl_2$ liquid does not occur.

For an explanation we turn to bond valence theory. The fact that in the crystalline form $Fe^{+3}Cl_3$ has an Fe-Cl bond length of ~2.27Å compared to ~2.51Å in $Fe^{+2}Cl_2$ can be attributed to the higher positive oxidation state having a smaller ionic size, and the changes in charge affecting the bonding orbitals and in the number of valence electrons participating in bonding. Based on bond valence theory we calculate the Shannon[50] effective ionic radii for the liquid state of $FeCl_2$ and $FeCl_3$. However, the ionic radii for 4-fold coordinated Cl does not exist based on existing crystal structures. From our experimental results we obtain the experimental Fe-Cl bond lengths if we assume the effective ionic radii for 4-fold coordinate Cl to be 1.71 Å. Comparing this result to the bond valence theory prediction for 6-fold coordinate Fe by Cl we would need to have at least a 11% increase in the $Fe^{+3}$-Cl bond length compared to $Fe^{+2}$-Cl bond length, but we only observe 7%. This explains why both liquid $FeCl_3$ and $FeCl_2$ are 4-fold coordinated by Cl.

Even though both $FeCl_2$ and $FeCl_3$ liquids exhibited the same predominant tetrahedral Fe-Cl coordination motif, the way these tetrahedra connected to one another, and thus the resulting network topology, can differ substantially. Such differences in network topology and degree of polymerization were assessed in the MD simulations by analyzing chloride atoms that acted as bridges between iron centers. Chloride ions that simultaneously coordinated two Fe cations, thereby forming an Fe-Cl-Fe linkage, were defined as "bridging-Cl". Operationally, these bridging configurations were identified by detecting Cl atoms bonded to two Fe ions, where bonding was determined using a specified cutoff distance.

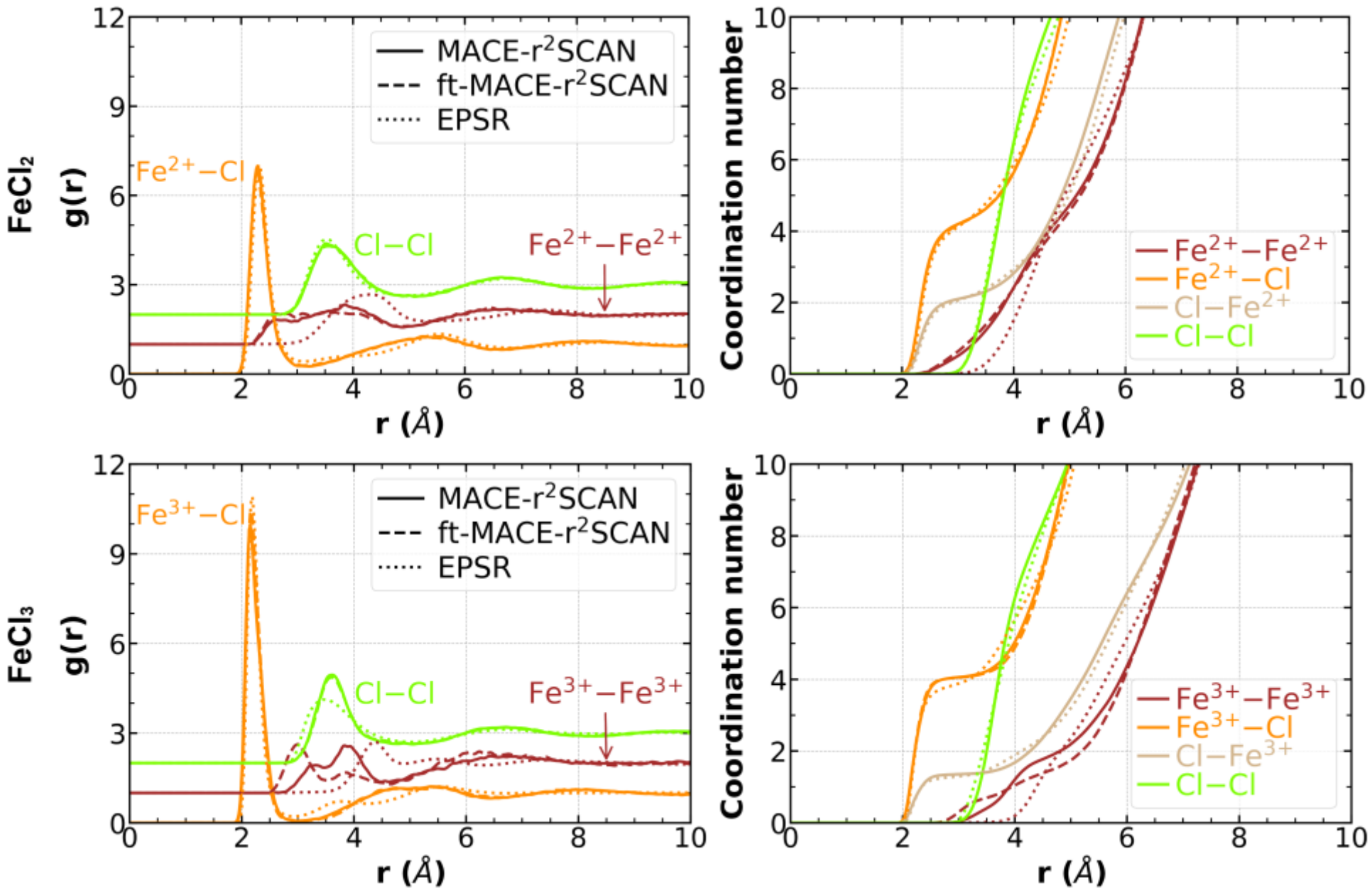

**Figure 5:** Fe-Cl, Cl-Cl and Fe-Fe partial RDFs (left column) and coordination numbers (right column) for molten $FeCl_2$ and $FeCl_3$ from MD simulations with MACE foundation model (MACE-$r^2$SCAN) and fine-tuned model (ft-MACE-$r^2$SCAN), and EPSR modelling.

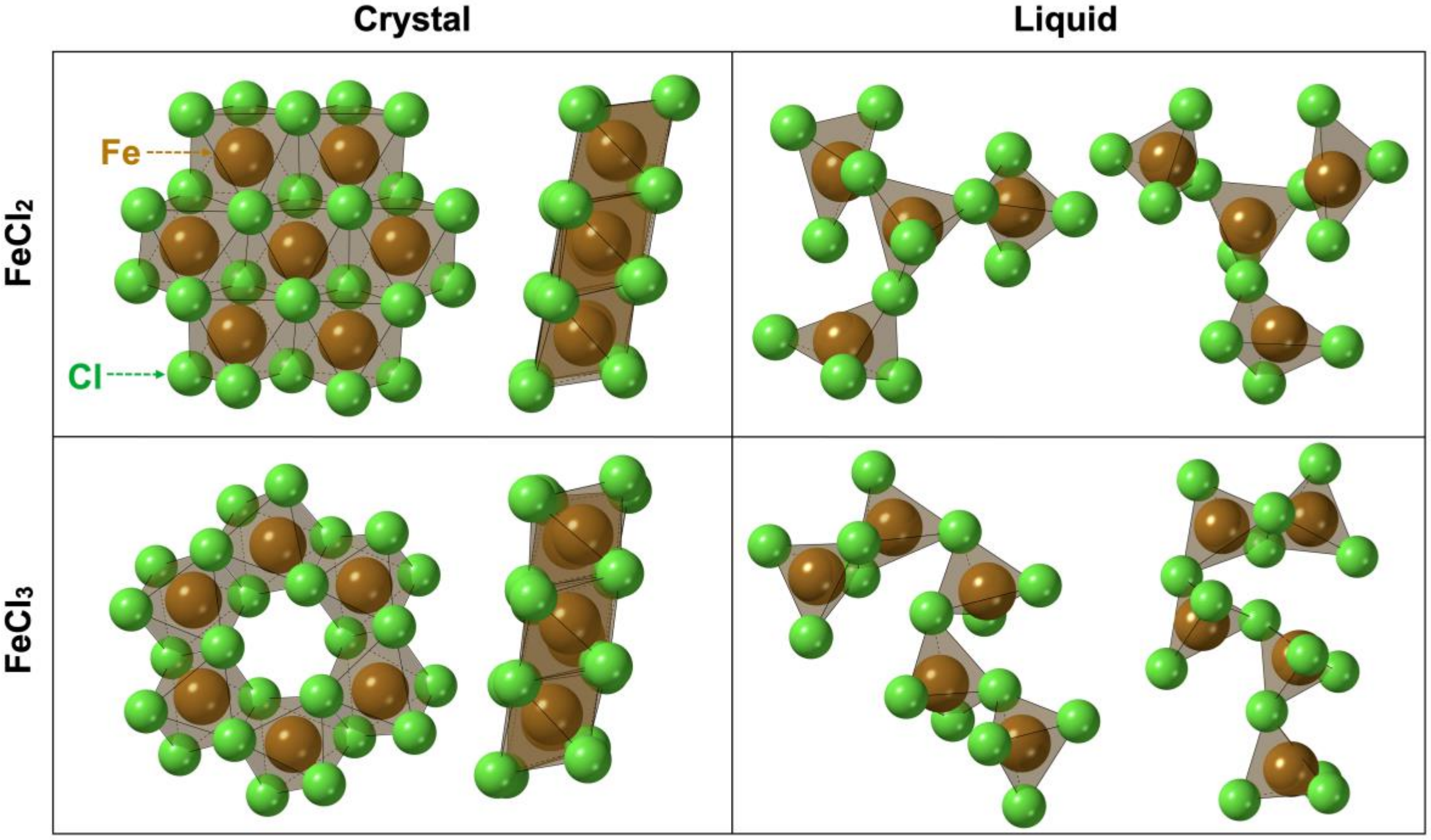

**Figure 6:** Representative configurations from the MACE-MD simulations illustrating the octahedral coordination in the crystals and the tetrahedral coordination in the corresponding melts for $FeCl_2$ and $FeCl_3$. Each of the four panels shows the front and the side view of the same group of atoms randomly selected from the corresponding simulation trajectory.

The resulting bridging-Cl fraction, evaluated as a function of Fe-Cl bond cutoff distance, is shown in Figure 9. Note, for each cutoff distance, the amount of bridging-Cl remained constant throughout the simulation; example shown in Figure S5 for 2.5 Å cutoff. In $FeCl_2$ the bridging-Cl fraction increased sharply, reaching ≈ 60 % at a cutoff of 2.5 Å and exceeding 95% by 3.5 Å. In contrast, the $FeCl_3$ melt exhibited a much more gradual rise, attaining only ~25% at 2.5 Å and ~55% at 4.0 Å before approaching saturation at larger cutoffs. The steep early increase for $FeCl_2$ indicated an extensively polymerized network of edge-sharing $FeCl_4$ tetrahedra in which most chloride ions participated in Fe-Cl-Fe linkages. The slower growth for $FeCl_3$ revealed a higher population of terminal (non-bridging) chlorides and correspondingly shorter oligomer chains.

The tetrahedral geometry and the connectivity patterns in the two melts were further evaluated by analyzing the Cl-Fe-Cl (intra-tetrahedral) and Fe-Cl-Fe (bridging) bond angle distributions extracted from the MACE-r$^2$SCAN MD trajectories (Figure 10). Here, A cutoff of 3 Å was used to define an Fe-Cl bond, chosen from the first minimum in the Fe-Cl RDF (Figure 5) and the plateau region observed in the bridging-chloride analysis (Figure 9). For both melts, the average Cl-Fe-Cl angle fell within 1.5° of the ideal tetrahedral value of 109.5°, confirming the predominantly tetrahedral coordination. However, notable differences emerged in the distribution shapes. In $FeCl_2$, the Cl-Fe-Cl distribution was right-skewed with a peak at ~100° and a broad tail extending to larger angles (standard deviation = 21.5°). In contrast, $FeCl_3$ exhibited a narrower, more symmetric distribution (standard deviation = 13.3°) centered closer to the ideal tetrahedral angle. This eight-degree difference in width indicates that $FeCl_4$ tetrahedra in $FeCl_2$ undergo greater angular distortion to accommodate the extended polymeric network, whereas the more isolated tetrahedra in $FeCl_3$ maintains closer-to-ideal geometry.

The Fe-Cl-Fe bridging-angle distributions revealed even more striking differences. $FeCl_2$ displayed a single, broad peak centered at ~103° with slight right-skewing, consistent with a relatively uniform edge-sharing motif throughout the polymerized network. Remarkably, $FeCl_3$ exhibited a bimodal distribution with peaks at ~87° and ~110°, despite having a narrower overall spread (9° smaller standard deviation than $FeCl_2$). This bimodality suggests two distinct bridging configurations in the $FeCl_3$ melt. Closer analysis of $Fe_2Cl_6$ units and their Fe-Cl-Fe bond angles in the $FeCl_3$ melt showed that the ~87° peak arose from these dimer-like species (Figure S6). In contrast, the ~110° peak corresponds to a more open Fe-Cl-Fe geometry and must be associated with larger oligomeric or network-like chloride-bridged Fe species. Taken together, these results indicate that molten $FeCl_3$ is structurally heterogeneous at the intermediate-range scale and explain the broad first peak observed in the Fe-Fe partial RDF (Figure 5) in the MACE-MD simulations of $FeCl_3$.

The differences in polymeric network structure and density between the two melts have direct implications for transport properties relevant to practical applications. Figure 11 presents the self-diffusion coefficients of $FeCl_2$ and $FeCl_3$ at three temperatures obtained from MD simulations using MACE-r$^2$SCAN. The calculated diffusion coefficients range from approximately $1.7 \times 10^{-5}$ to $5.5 \times 10^{-5}$ cm$^2$/s, consistent with experimentally reported values for molten salts in this temperature regime, which typically fall within $10^{-5}$ to $10^{-4}$ cm$^2$/s. Notably, $FeCl_3$, characterized by a mix of both short and intermediate-sized polyhedral chains and lower density, attained diffusion coefficients comparable to those of $FeCl_2$ at temperatures nearly 200 K lower. Consequently, the activation energy ($E_a$) for ionic mobility for $FeCl_3$ was lower (0.18 eV) than for $FeCl_2$ (0.32 eV). This enhanced mobility reflects the less interconnected network structure and reduced packing constraints for $FeCl_3$. Overall, these results demonstrate how differences in atomic-scale connectivity translate directly into measurable transport behavior, highlighting the importance of accurate structural models for predicting and optimizing molten salt performance in energy applications.

## Conclusion

This combined HEXRD, EPSR, and MACE-MD investigation clarified the liquid structures of $FeCl_2$ and $FeCl_3$ and provided a revised structural picture for molten iron chlorides. HEXRD showed clear differences in both short-range and intermediate-range order between the two melts, with the first RDF peak appearing at 2.33 Å for $FeCl_2$ and 2.19 Å for $FeCl_3$. MD simulations using the three different MACE foundation models, namely, MACE-MPA-0, MACE-MATPES-PBE-0, and MACE-MATPES-r$^2$SCAN-0, were performed to investigate the molten structures closely. Among the three MACE models, the MACE-r$^2$SCAN foundation model performed the best in reproducing experimental structure factors and RDFs. In addition, EPSR analysis was carried out to complement MACE-calculated structure factors and RDFs.

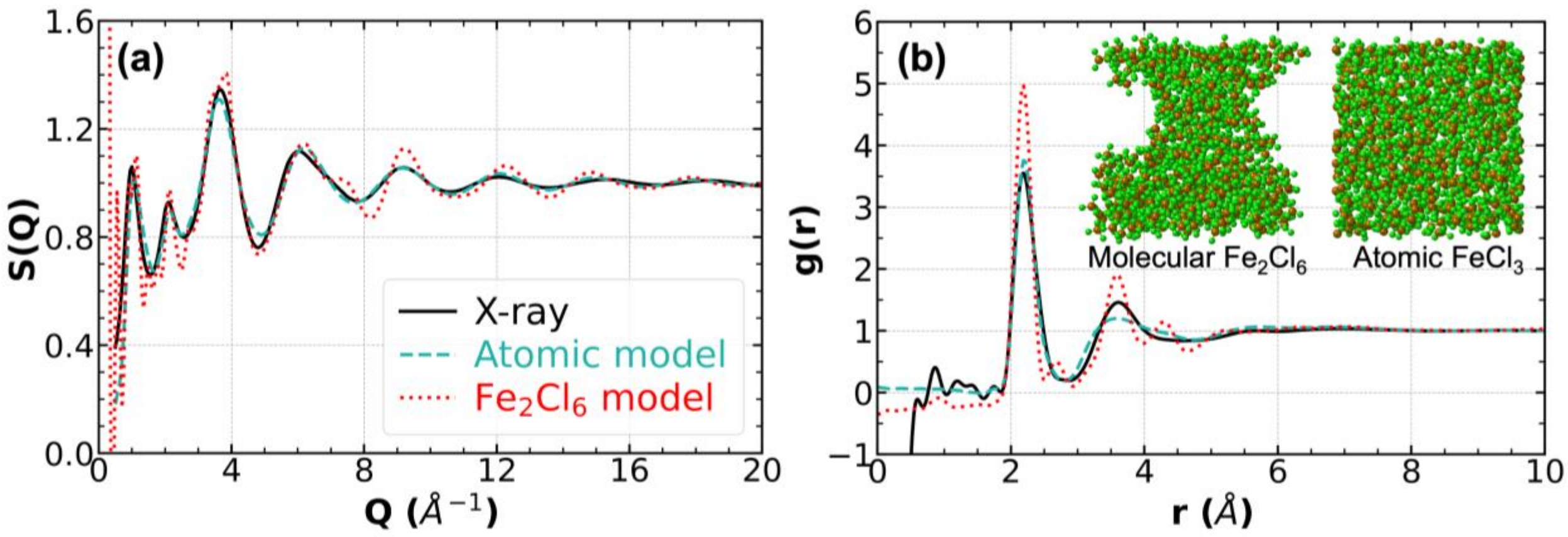

**Figure 7:** (a) The measured x-ray structure factor of molten $FeCl_3$ at 310C (black circles) compared to our EPSR $Fe_2Cl_6$ molecular model (solid red line) and atomic $FeCl_3$ model (blues dashed line). The insert shows the arrangement of molecules and atoms in the EPSR simulation box for the two models fixed at the experimental density. (b) the corresponding x-ray pair distribution function obtained from a Sine Fourier transform of the data in (a) together with a schematic of the $Fe_2Cl_6$ molecule observed in the vapor phase.

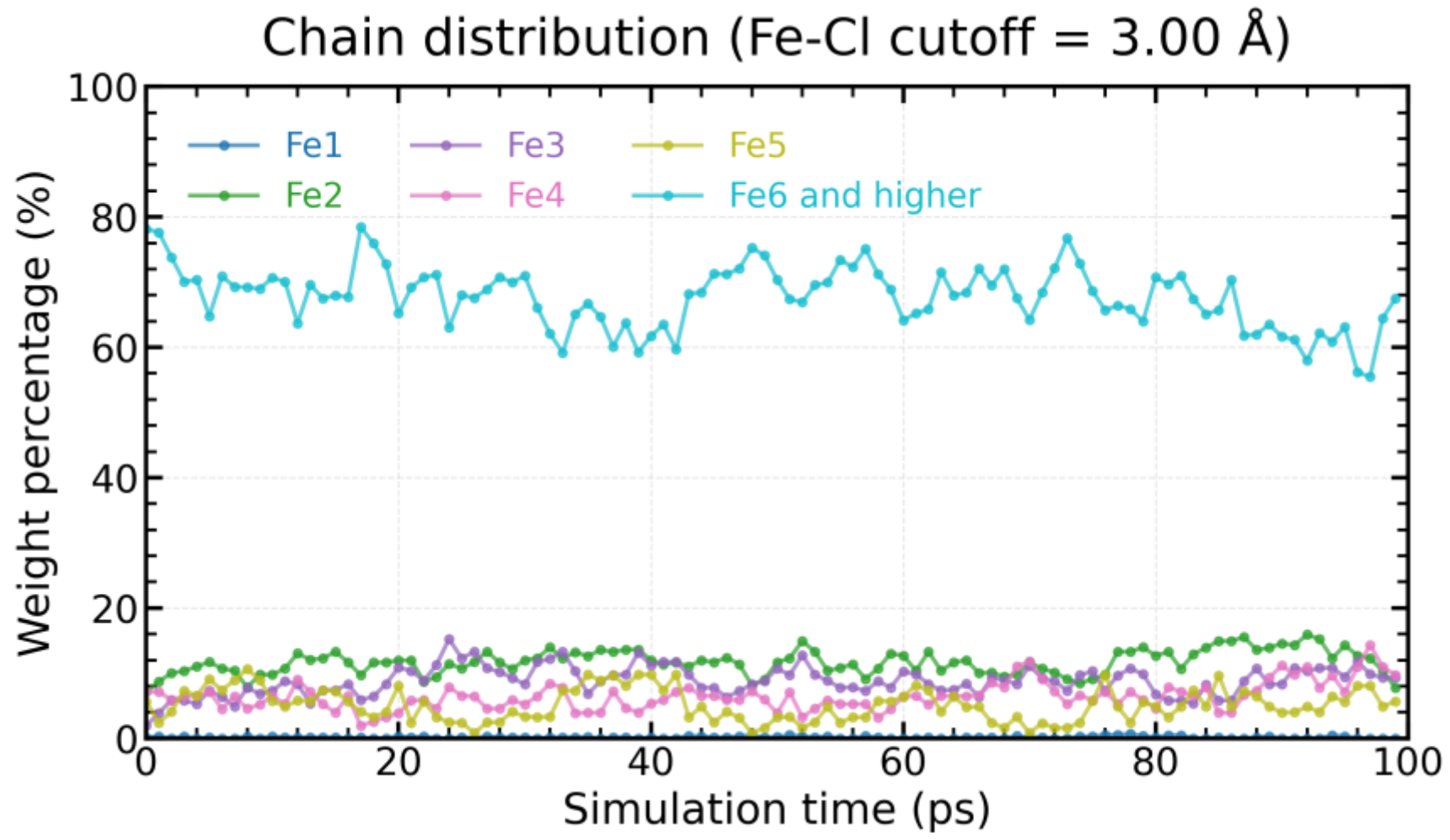

**Figure 8:** Analysis of $Fe_xCl_y$ polyhedral chain weights from MD simulations of liquid $FeCl_3$ performed using the MACE-r$^2$SCAN foundation model. Polymerized chains are categorized by the number of Fe ions they contain, and the total mass of each category is normalized by the total simulation box mass to report weight percentages. Fe-Cl cutoff used was 3.0 Å for detecting the polyhedral chains.

The MACE-$r^2$SCAN model, together with EPSR analysis, reproduced the experimental structure factors and total RDFs with generally good agreement, although some differences remained in the low-Q region and in Fe-Fe correlations. The MACE-$r^2$SCAN model was fine-tuned using $FeCl_2$ crystal and liquid structures to further minimize the small discrepancies at low-Q region. The fine-tuned model performance followed the foundation model closely with marginal improvements, suggesting that a higher level of DFT theory than $r^2$SCAN may be required to generate more accurate training data for molten iron salts.

Notably, MACE-MD simulations correctly captured the octahedral coordination of Fe in crystal $FeCl_3$ and the tetrahedral coordination of Fe in liquid $FeCl_3$. Further, the MACE models predicted that $FeCl_2$ also undergoes a similar coordination change upon melting, from octahedral $FeCl_6$ environments in the crystalline phases to predominantly tetrahedral $FeCl_4$ environments in the liquids accompanied by a 26% decrease in density. The present results therefore showed that $FeCl_2$, like $FeCl_3$, does not retain six-fold Fe coordination in the melt.

Closer analysis of the local structures of the two melts and the Fe-Cl bonding patterns in the MD simulations showed that similar local coordination does not imply similar liquid structure. Molten $FeCl_2$ in the MD simulations formed an extensively polymerized network of edge-sharing $FeCl_4$ tetrahedra, with the bridging-chloride fraction increasing rapidly to about 60% at a 2.5 Å Fe-Cl cutoff and exceeding 95% by 3.5 Å cutoff. In contrast, molten $FeCl_3$ in the MD simulations exhibited a more diverse oligomeric network with a much larger population of terminal chlorides, reaching only about 25% bridging at 2.5 Å and about 55% even at 4.0 Å. These differences were further reflected in the angular distributions. $FeCl_2$ showed a broad and distorted Cl-Fe-Cl distribution and a single broad Fe-Cl-Fe bridging-angle peak near 103°, consistent with a flexible, continuously connected polymeric network. $FeCl_3$ instead exhibited a narrower Cl-Fe-Cl distribution together with a bimodal Fe-Cl-Fe distribution, with peaks near 87° and 110°, indicating two distinct bridging motifs. Closer analysis of the $FeCl_3$ melt associated the peak near 87° to $Fe_2Cl_6$ molecules.

A key outcome of the present study is a reassessment of the accepted liquid-state model of $FeCl_3$. EPSR modeling showed that a liquid composed primarily of discrete $Fe_2Cl_6$ molecules could not simultaneously reproduce the HEXRD data and the experimental density, whereas an atomic model based on connected $FeCl_4$ units reproduced both. Analysis of the length of the Fe-Cl polyhedral chains in the MACE-MD simulations supported the same conclusion: $Fe_2Cl_6$ units accounted for only a minor fraction of the melt, whereas extended chains containing six or more Fe centers dominated the structure. Molten $FeCl_3$ should therefore not be described primarily as a molecular liquid of isolated $Fe_2Cl_6$ dimers, but rather as a liquid dominated by short-to-extended polyhedral chain networks.

Finally, the structural differences between the two melts were directly linked to transport behavior by calculating the diffusion coefficients. The less connected and lower-density $FeCl_3$ melt exhibited diffusion coefficients comparable to those of $FeCl_2$ at temperatures nearly 200 K lower, indicating higher mobility. These atomistic insights provide a quantitative framework for understanding transport properties and electrochemical behavior in iron chloride melts, with direct implications for optimizing molten-salt electrolytes in iron production and redox flow batteries. More broadly, this work demonstrates the efficacy of combining scattering experiments with data-driven interatomic potentials to resolve complex liquid structures in technologically relevant molten salts.

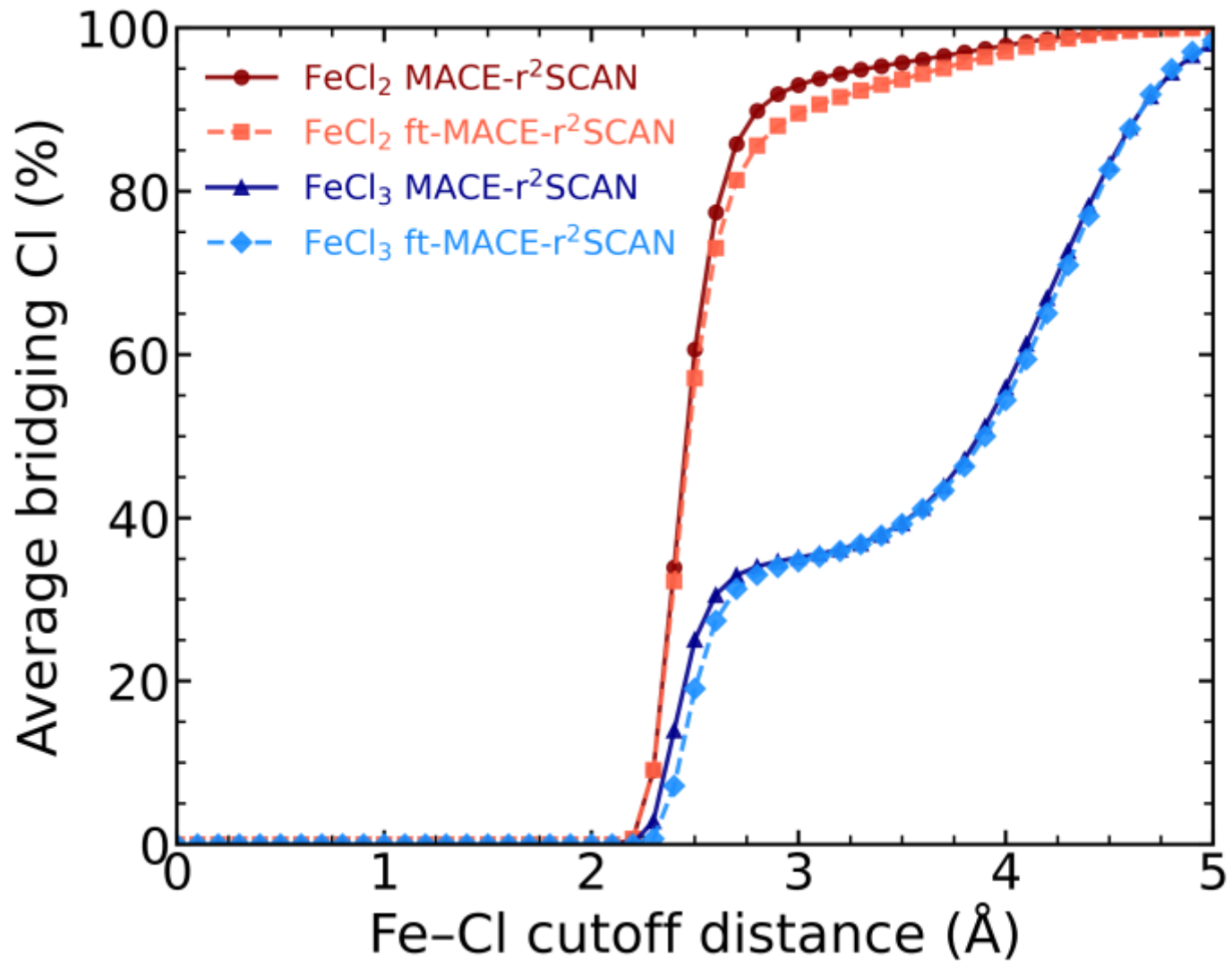

**Figure 9:** Variation of the fraction of bridging-Cl atoms, i.e., chlorides acting as bridges between iron atoms, as a function of the Fe-Cl cutoff distance in $FeCl_2$ and $FeCl_3$ melts in the MD simulations performed using the MACE-r$^2$SCAN foundation model and the fine-tuned model (ft-MACE-r$^2$SCAN).

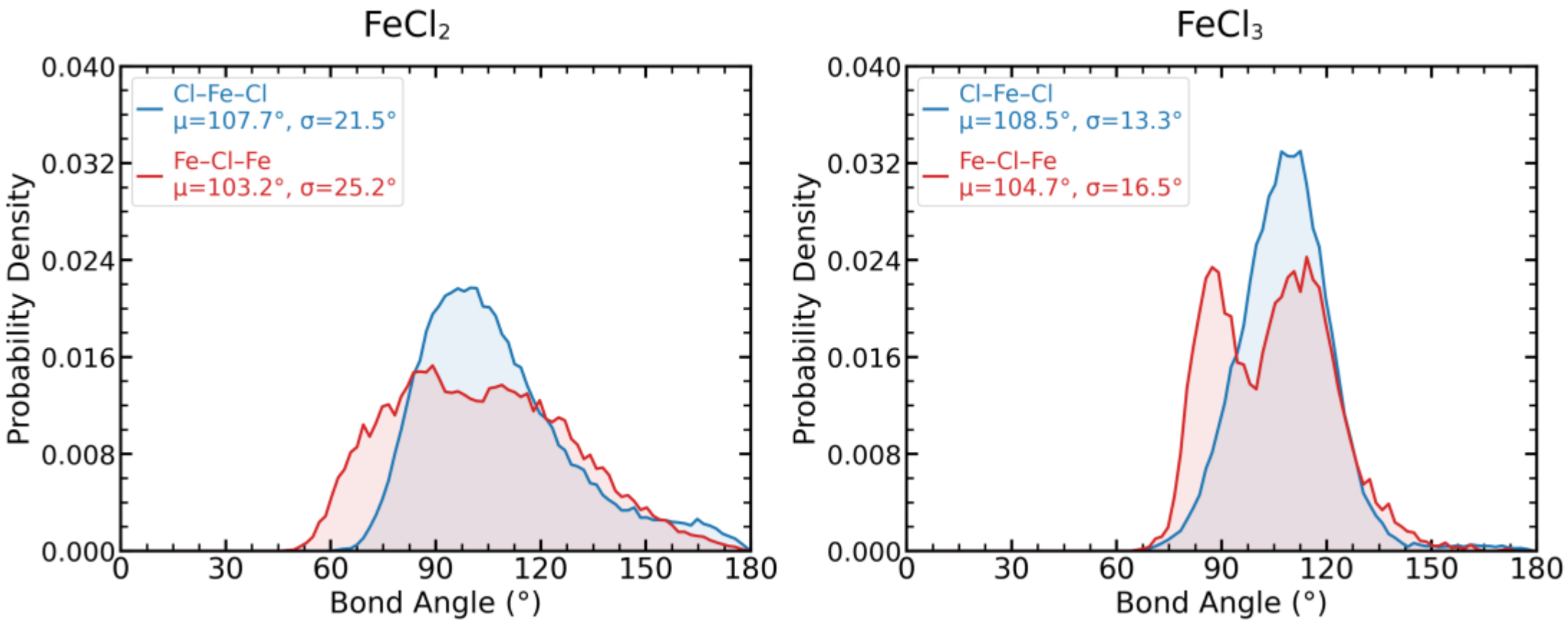

**Figure 10:** Distributions of Cl-Fe-Cl (intra-tetrahedral) and Fe-Cl-Fe (bridging) angles from MD simulations employing the MACE-r$^2$SCAN foundation model for $FeCl_2$ and $FeCl_3$. Fe-Cl bond

cutoff was 3.0 Å. The μ and the σ in the plot legends represent the average and the standard deviation, respectively, of the distributions.

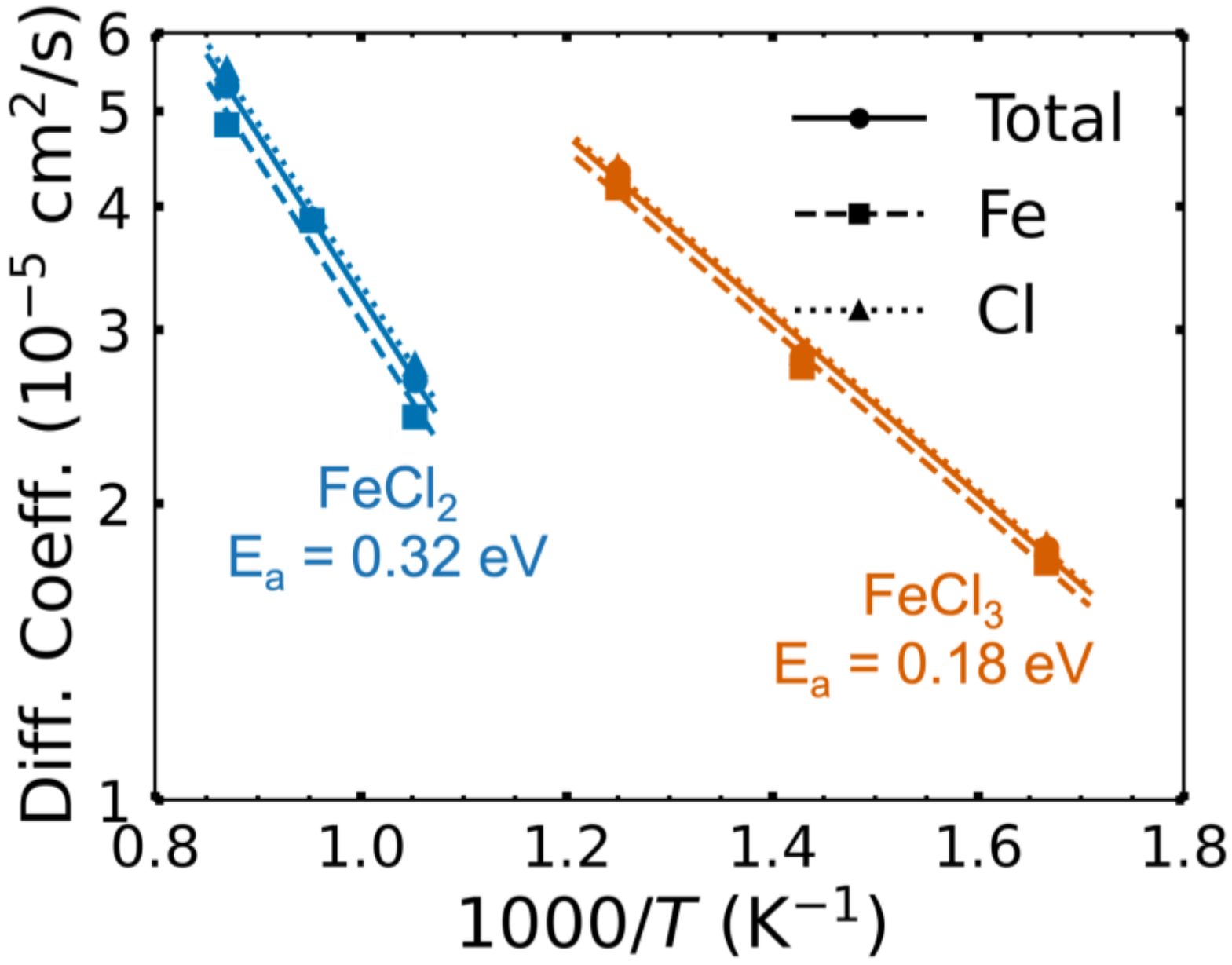

**Figure 11:** Arrhenius plot of the calculated diffusion coefficients for molten $FeCl_2$ and $FeCl_3$ from MD simulations employing the MACE-r$^2$SCAN foundation model. The total, Fe, and Cl diffusion coefficients are plotted on a logarithmic scale against inverse temperature (1000/T). Blue and orange markers denote the $FeCl_2$ and $FeCl_3$ systems, respectively. The straight lines represent linear least-squares fits to the Arrhenius equation ($D = D_0 \exp[-E_a/k_B T]$), from which the activation energies, $E_a$, for ionic mobility are derived.

## Acknowledgment

This material is based upon work supported by the U.S. Department of Energy, Office of Science Energy Earthshot Initiative as part of the Center for Steel Electrification by Electrosynthesis (C-STEEL) at Argonne National Laboratory under Contract Number DE-AC02-06CH11357. We acknowledge the computing resources provided on 'Improv' and 'Swing' computing clusters operated by the Laboratory Computing Resource Center at Argonne National Laboratory. This research used resources of the Argonne Leadership Computing Facility, which is a U.S. Department of Energy Office of Science User Facility operated under contract DE-AC02-06CH11357. A.V.M. and F.H.B. were supported by the Office of Science, U.S. Department of Energy, under contract DE-AC02-06CH11357.

## Author Contributions

F.H.B. wrote the manuscript, performed MLIP fine-tuning, carried out all MD simulations, and analyzed MD simulation data. J.G., C.J.B., A.B., and D.J. performed all HEXRD experiments and analyzed experimental data. J.G. wrote the experimental portion of the Method section. C.J.B. performed the EPSR calculations and analyzed EPSR data. C.J.B. wrote portion of the Discussion section. A.B. and D.J. reviewed the manuscript. F.H.B., J.G., C.J.B., and A.V.M edited and reviewed the manuscript. A.V.M. supervised the overall work and conceived the research plan.

## Competing Interest Statement

The authors declare no competing interest.

## Supplementary Information

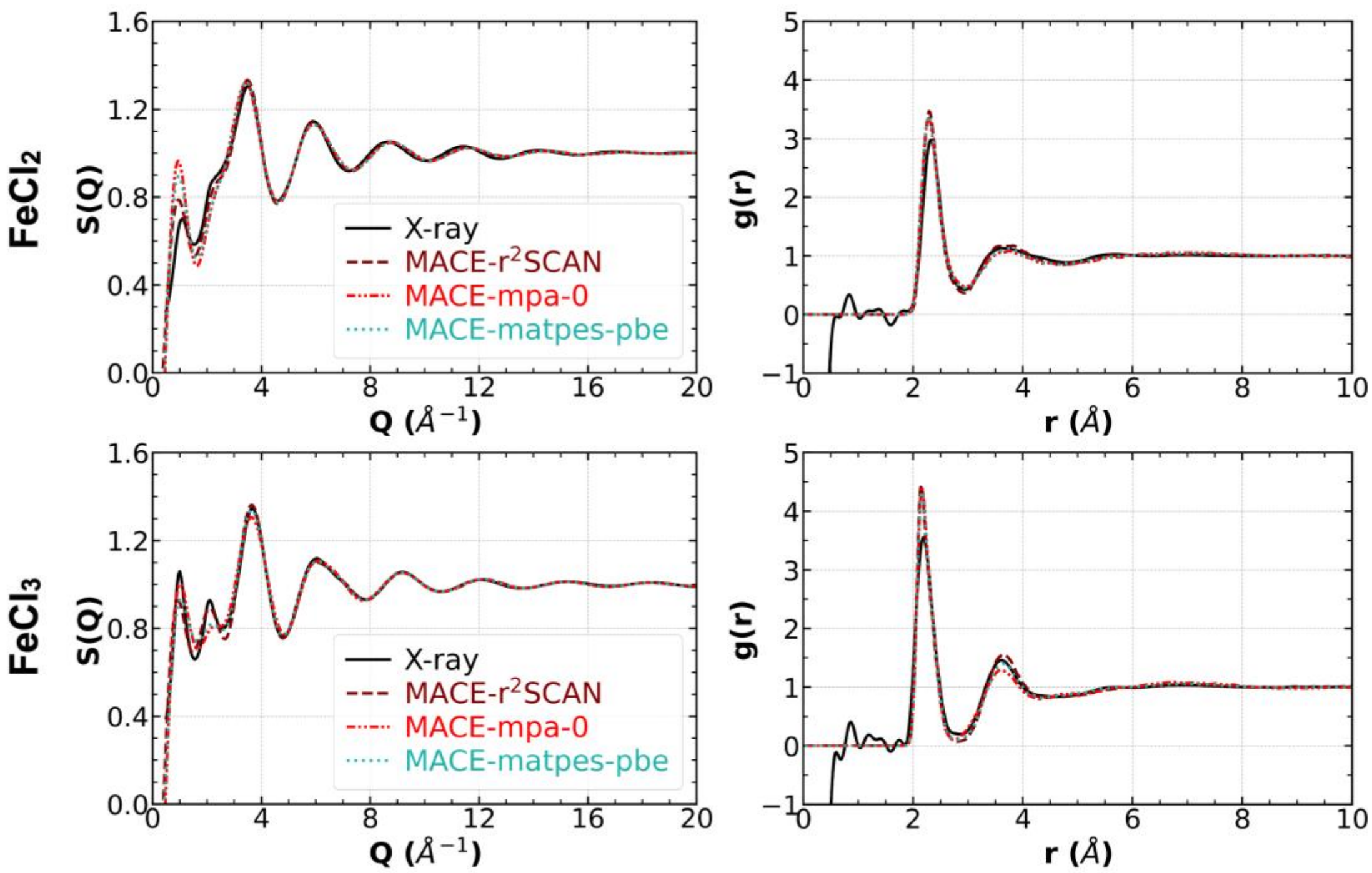

**Figure S1:** Comparison of MACE foundation model performance for modelling molten iron salts. Structure factors are shown on the left panel and RDFs on the right panel. Three foundation models, i.e., MACE-MPA-0, MACE-MATPES-PBE-0, and MACE-MATPES-r$^2$SCAN-0 were tested and compared to HEXRD (X-ray) experimental results.

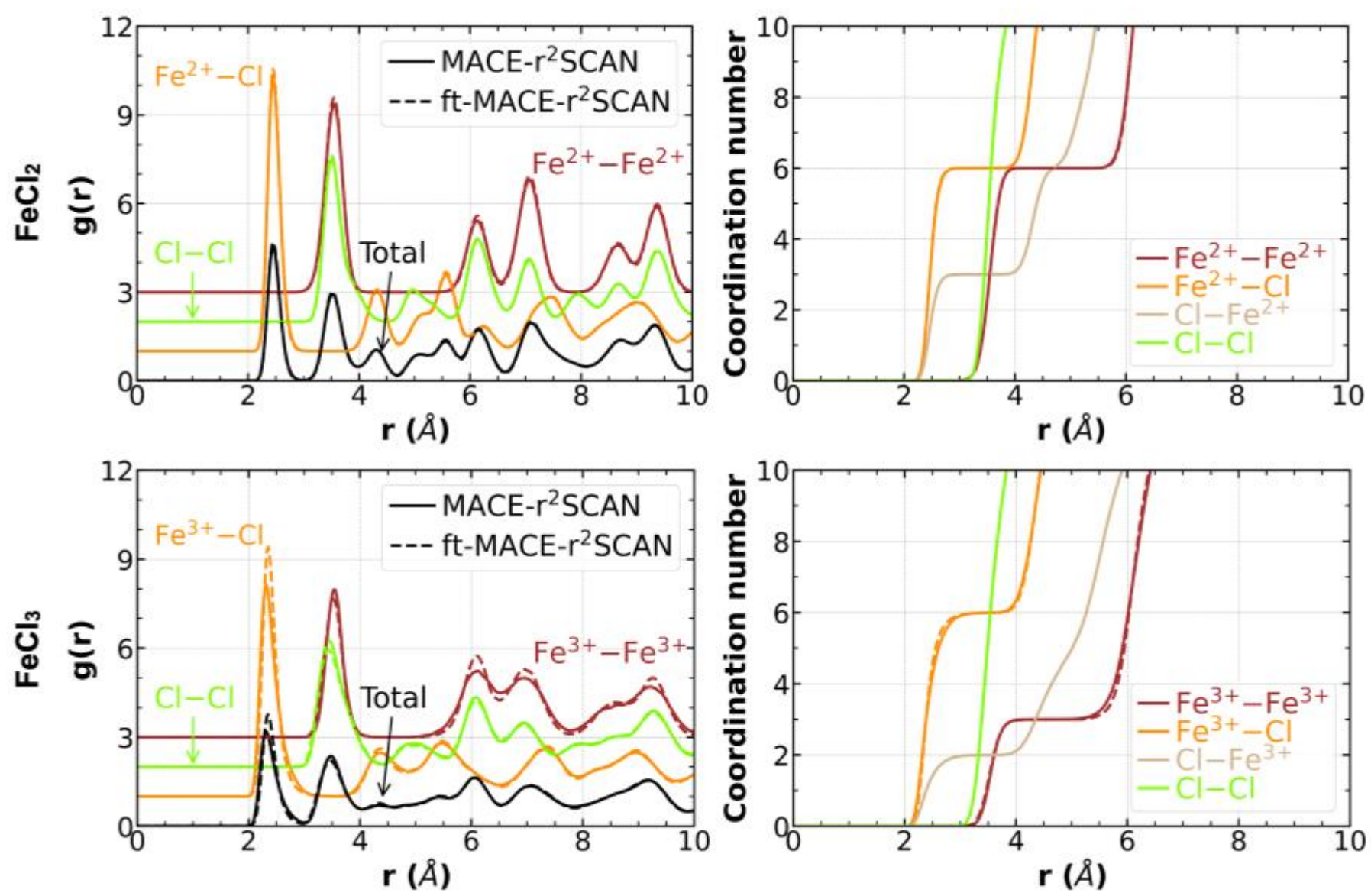

**Figure S2:** Fe–Cl, Cl–Cl and Fe–Fe partial RDFs (left column) and coordination numbers (right column) for crystal $FeCl_2$ and $FeCl_3$ from MD simulations with MACE foundation model (MACE-r$^2$SCAN) and fine-tuned model (ft-MACE-r$^2$SCAN).

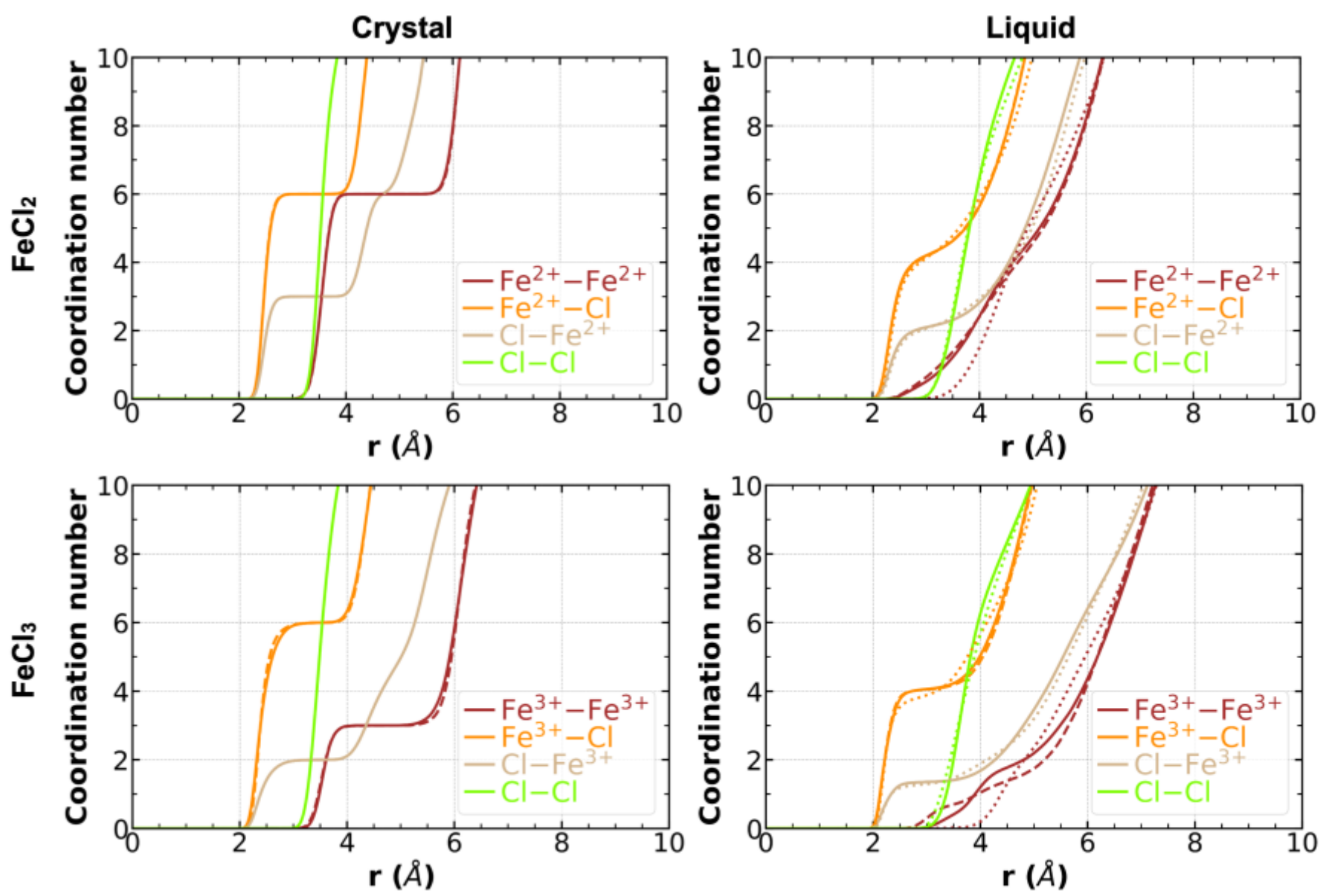

**Figure S3:** Comparison of the coordination numbers in crystal (left panel) vs liquid (right panel) phase calculated from MD simulations using MACE-r$^2$SCAN foundation model (solid lines) and the fine-tuned model (dotted lines).

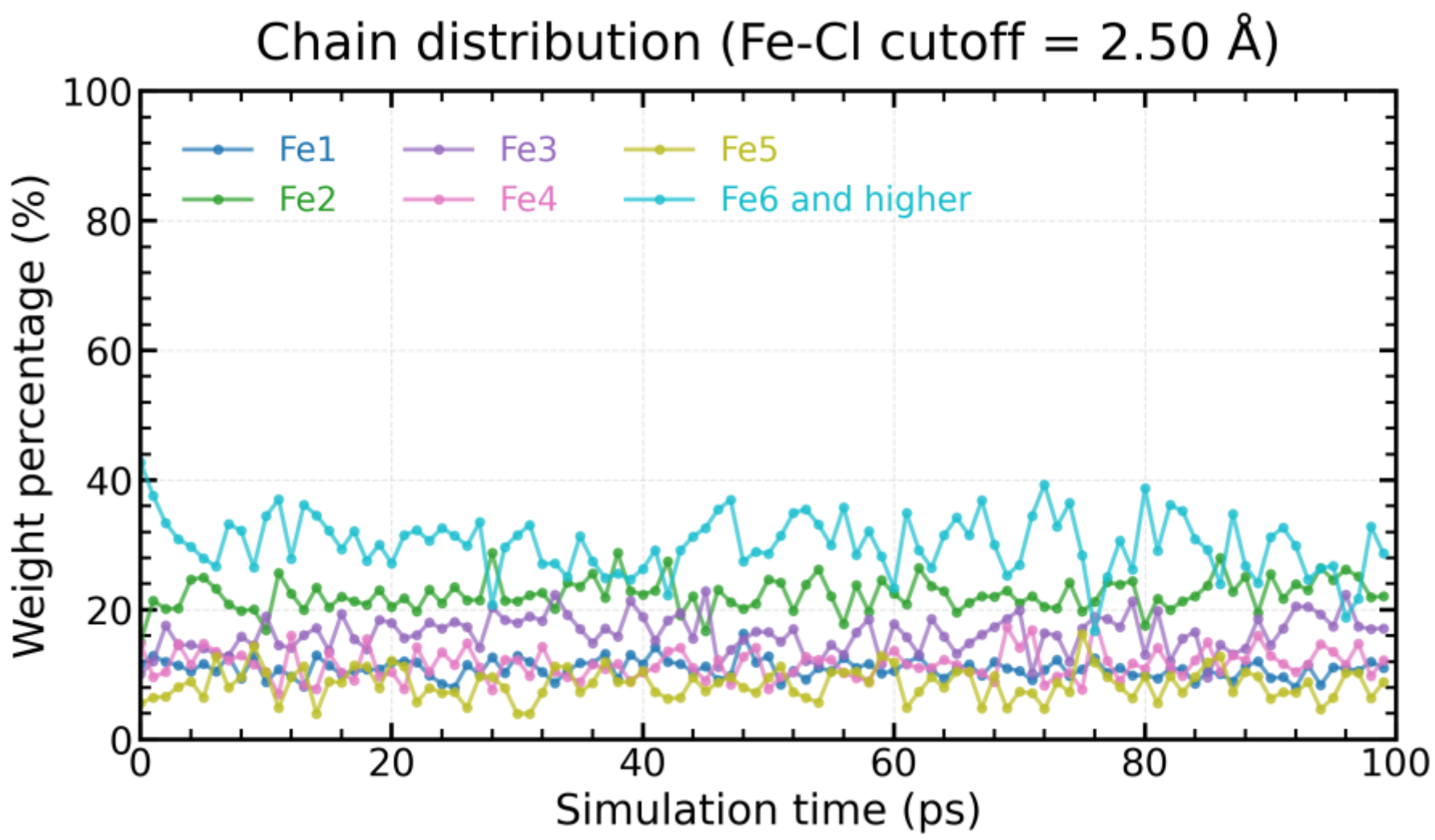

**Figure S4:** Analysis of $Fe_xCl_y$ polyhedral chain weights from MD simulations of liquid $FeCl_3$ performed using the MACE-r$^2$SCAN foundation model. Polymerized chains are categorized by the number of Fe ions they contain, and the total mass of each category is normalized by the total simulation box mass to report weight percentages. Fe-Cl cutoff used was 2.5 Å for detecting the polyhedral chains.

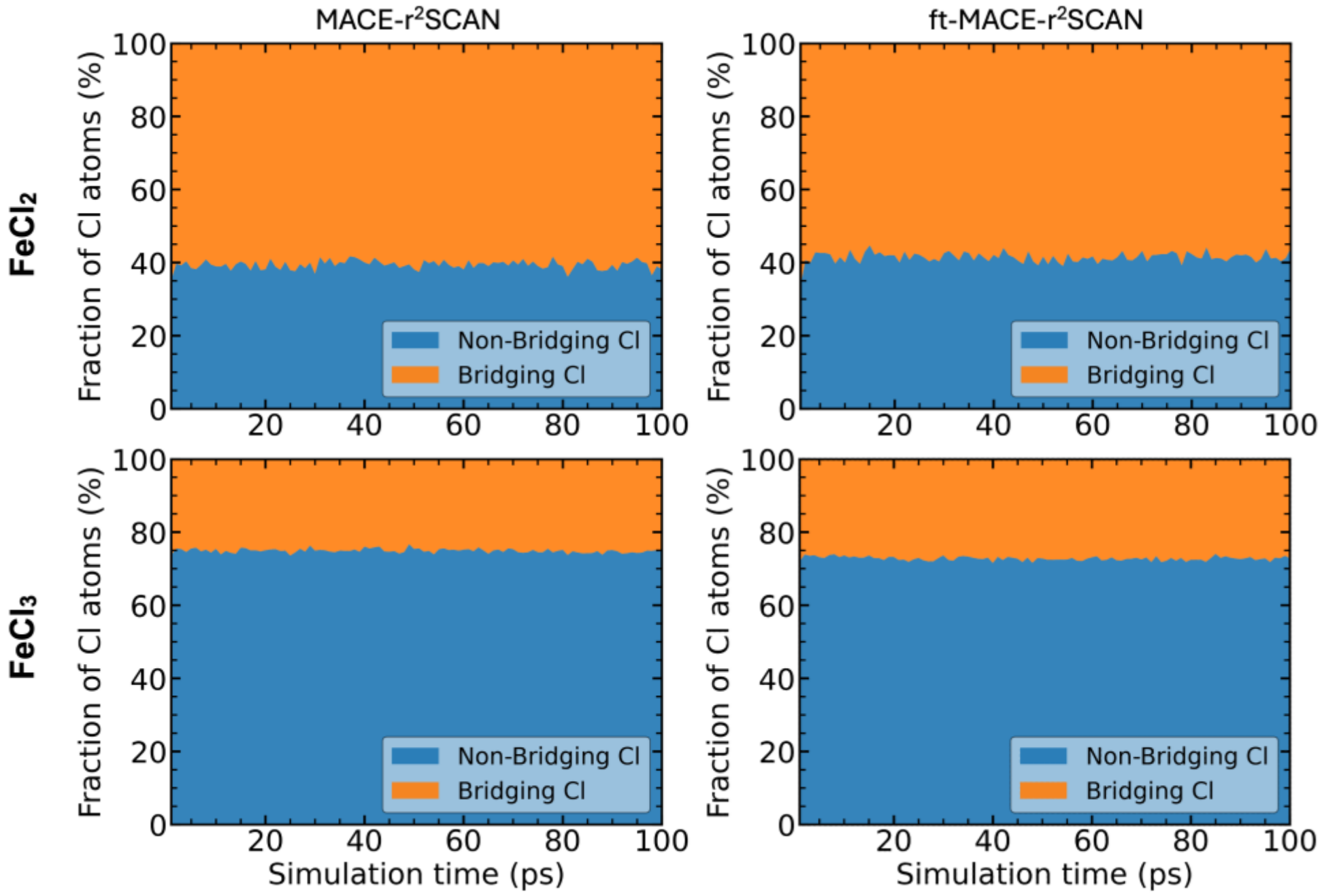

**Figure S5:** Variation of the fraction of bridging Cl in the MACE-MD simulations. Fe-Cl cutoff used was 2.5 Å for detecting bonds.

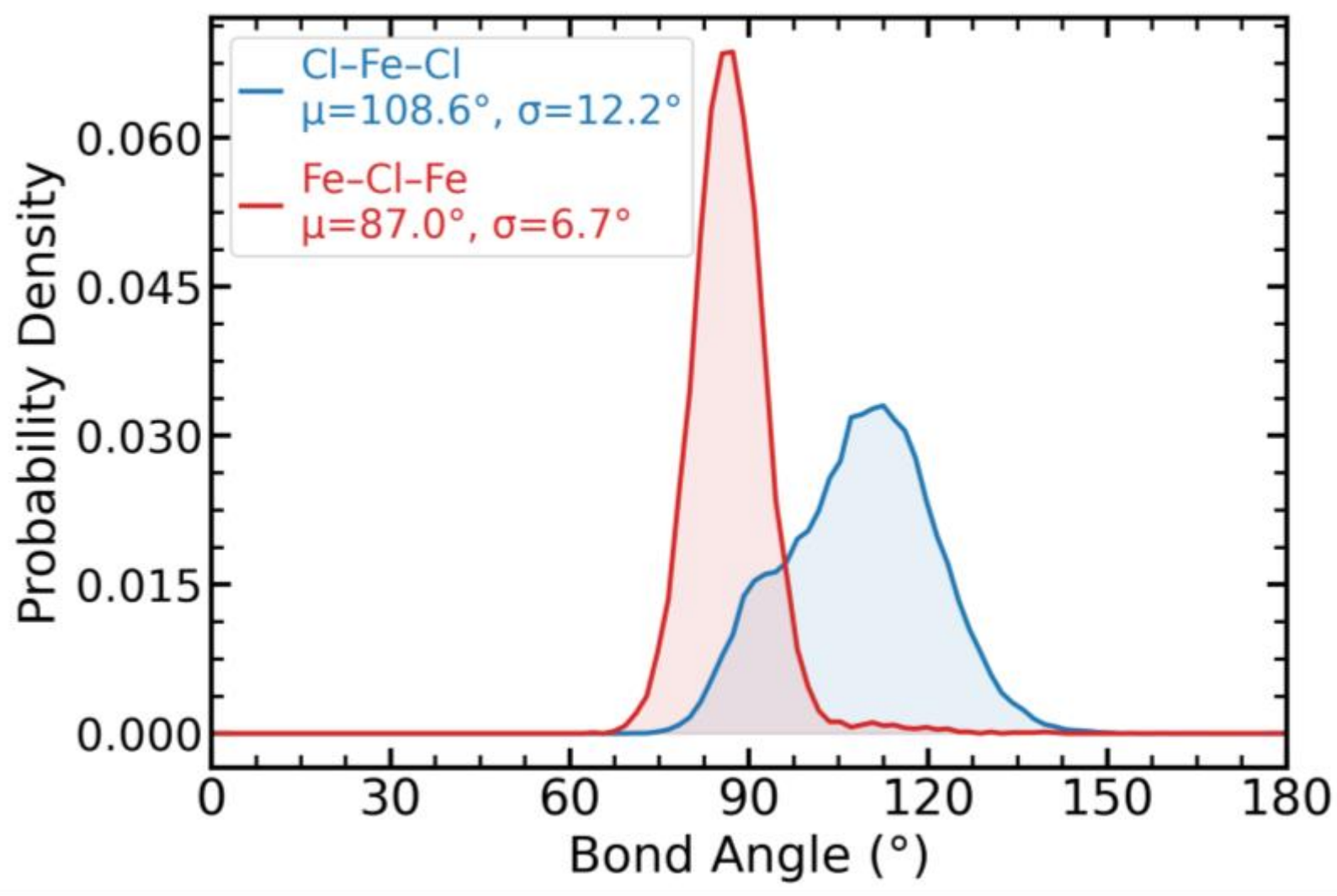

**Figure S6:** Distributions of Cl-Fe-Cl (intra-tetrahedral) and Fe-Cl-Fe (bridging) angles of only the $Fe_2Cl_6$ molecules from MD simulations employing the MACE-$r^2$SCAN foundation model for $FeCl_3$. Fe-Cl bond cutoff was 3.0 Å. The μ and the σ in the plot legends represent the average and the standard deviation, respectively, of the distributions.